\documentclass[doublespacing]{elsart}

\usepackage{graphicx}
\usepackage{amssymb}
\usepackage{amsmath}

\begin{document}

\begin{frontmatter}

% Title, authors and addresses

% use the thanksref command within \title, \author or \address for footnotes;
% use the corauthref command within \author for corresponding author footnotes;
% use the ead command for the email address,
% and the form \ead[url] for the home page:
% \title{Title\thanksref{label1}}
% \thanks[label1]{}
% \author{Name\corauthref{cor1}\thanksref{label2}}
% \ead{email address}
% \ead[url]{home page}
% \thanks[label2]{}
% \corauth[cor1]{}
% \address{Address\thanksref{label3}}
% \thanks[label3]{}

\title{$2\alpha+t$ cluster structure in $^{11}$B}

% use optional labels to link authors explicitly to addresses:
% \author[label1,label2]{}
% \address[label1]{}
% \address[label2]{}

\author[CNST]{T.~Kawabata\corauthref{cor1}},
\ead{kawabata@cns.s.u-tokyo.ac.jp}
\corauth[cor1]{Corresponding author.}
\author[KONA]{H.~Akimune},
\author[WITW]{H.~Fujita},
\author[OSAK]{Y.~Fujita},
\author[RCNP,JAEA]{M.~Fujiwara},
\author[KEK]{K.~Hara},
\author[RCNP]{K.~Hatanaka},
\author[CYRI]{M.~Itoh},
\author[YITP]{Y.~Kanada-En'yo},
\author[KYTO]{S.~Kishi},
\author[CNST]{K.~Nakanishi},
\author[MIYA]{H.~Sakaguchi},
\author[MSU]{Y.~Shimbara},
\author[RCNP]{A.~Tamii},
\author[RIKE]{S.~Terashima},
\author[TITP]{M.~Uchida},
\author[KYUS]{T.~Wakasa},
\author[TSUK]{Y.~Yasuda},
\author[KYUY]{H.~P.~Yoshida},
\author[RCNP]{M.~Yosoi}

\address[CNST]{Center for Nuclear Study, Graduate School of Science,
University of Tokyo, Saitama 351-0198, Japan}
\address[KONA]{Department of Physics, Konan University,
Kobe, Hyogo 658-8501, Japan}
\address[WITW]{School of Physics, University of the Witwatersrand,
Johannesburg, 2050, South Africa}
\address[OSAK]{Department of Physics, Osaka University,
Toyonaka, Osaka 560-0043, Japan}
\address[RCNP]{Research Center for Nuclear Physics, 
Osaka University, Ibaraki, Osaka 567-0047, Japan}
\address[JAEA]{Kansai Photon Science Institute, 
Japan Atomic Energy Agency, Kizu, Kyoto 619-0215, Japan}
\address[KEK]{KEK, High Energy Accelerator Research Organization,
Tsukuba, Ibaraki 305-0801, Japan}
\address[CYRI]{
Cyclotron and Radioisotope Center (CYRIC), Tohoku University,
Sendai, Miyagi 980-8578, Japan}
\address[YITP]{Yukawa Institute for Theoretical Physics, Kyoto University,
Kyoto 606-8502, Japan}
\address[KYTO]{Department of Physics, Kyoto University,
Kyoto 606-8502, Japan}
\address[MIYA]{Faculty of Engineering, Miyazaki University, Miyazaki
889-2192, Japan}
\address[MSU]{National Superconducting Cyclotron Laboratory, 
Michigan State University, East Lansing, Michigan 48824, USA}
\address[RIKE]{RIKEN (The Institute for Physical and Chemical Research),
Wako, Saitama 351-0198, Japan}
\address[TITP]{Department of Physics, Tokyo Institute of Technology,
Meguro, Tokyo 152-8551, Japan}
\address[KYUS]{Department of Physics, Kyushu University, 
Fukuoka 812-8581, Japan}
\address[TSUK]{Institute of Physics, University of Tsukuba, Ibaraki
305-8571, Japan}
\address[KYUY]{
Research and Development Center for Higher Education,Kyushu University,
 Fukuoka 810-8560, Japan}

\begin{abstract}
The cluster structures of the excited states in $^{11}$B are
studied by analyzing the isoscalar monopole and quadrupole strengths in
the $^{11}$B($d$,$d'$) reaction at $E_d=200$~MeV.
The excitation strengths are compared with the predictions
by the shell-model and antisymmetrized molecular-dynamics (AMD)
calculations. 
The large monopole strength for
the $3/2^-_3$ state at $E_x=8.56$~MeV is well described by the AMD
calculation and is suggested to be an evidence for a well developed
$2\alpha+t$ cluster structure. 
\end{abstract}

\end{frontmatter}

% main text
%\section{}
%\label{}

%%%%%%%%%%%%%%%%%%%%%%%%%%%%%%%%%%%%%%%%%%%%%%%%%%%%%%%%%%%%%%%%%%%%%%%
%\section{Introduction}

Alpha particle clustering is an important concept in nuclear physics
for light nuclei.
On the basis of the Ikeda diagram \cite{IKED68}, the cluster structure
is expected to emerge near the $\alpha$-decay threshold energy.
It has been suggested
that the 7.65-MeV $0^+_2$ state in $^{12}$C, which locates at
an excitation energy higher than the 3$\alpha$-decay threshold
by 0.39~MeV, has a 3$\alpha$-cluster configuration
\cite{MORI56,FUJI80,UEGA77,UEGA79,KAMI81}. 
This $0^+_2$ state is theoretically described by
introducing a novel concept of the nuclear structure, {\it i.e.},
this state has a dilute-gas-like structure where three $\alpha$
particles are weakly interacting and are condensed into the lowest
$s$-orbit \cite{TOHS01,FUNA03,FUNA05,YAMA05,MATS04}. 
It has been found that the wave function of the $0^+_2$ state calculated
by the previous 3$\alpha$ cluster models \cite{UEGA77,UEGA79,KAMI81} is 
almost equivalent to the wave function of the $3\alpha$ condensed state
\cite{FUNA03}.
Similar dilute-gas states of $\alpha$
clusters have been predicted in self-conjugate $N=4n$ nuclei \cite{YAMA04}.
The next natural question addressed is whether such a dilute cluster
state exists in the other $N\,{\ne}\,4n$ nuclei such as $^{11}$B.

%Recently, an exotic character of the $3/2^-_3$ state at $E_x=$ 8.56~MeV
%in $^{11}$B was found in the measurement of the Gamow-Teller (GT) and
%spin-flip $M1$ strengths for excited states in $^{11}$B and its
%analog in $^{11}$C \cite{PAPA04,FUJI04,HANN03}. The GT and spin-flip $M1$
%strengths for the $3/2^-_3$ state are abnormally quenched in comparison
%with the other states. 
%
%The $3/2^-_3$ state in $^{11}$B locates at the excitation energy lower
%than the $\alpha$-decay threshold by 100 keV and is not well described
%by the shell-model (SM) calculations. 
%Hence, it is very interesting to
%study the nuclear structure of this $3/2^-_3$ state from the perspective
%of clustering.

Recently, an exotic character of the $3/2^-_3$ state at $E_x=$ 8.56~MeV
in $^{11}$B was found in the measurement of the Gamow-Teller (GT) and
spin-flip $M1$ strengths for excited states in $^{11}$B and its
analog in $^{11}$C \cite{PAPA04,FUJI04,HANN03}. The GT and spin-flip $M1$
strengths for the $3/2^-_3$ state are abnormally quenched in comparison
with the other states. 

The abnormally quenched strengths imply that the structure of the
$3/2^-_3$ state is quite different from those of the other low-lying
states. The $3/2^-_3$ state is not well described by the shell-model
(SM) calculations while the other states including the ground state
are successfully described. It is naturally noted that the
$3/2^-_3$ state has a non-SM-like structure. The non-SM-like structure
of the $3/2^-_3$ state is possibly due to the $\alpha$ clustering.
Since the theoretical description of the clustering phenomena under the
SM framework requires a huge number of single-particle bases, it is
generally difficult to treat the clustering phenomena in the truncated
SM space.

The $3/2^-_3$ state in $^{11}$B actually locates at the excitation
energy lower than the $\alpha$-decay threshold by 100 keV where the
cluster structure is expected to emerge. Hence, it is very interesting
to study the nuclear structure of this $3/2^-_3$ state from the
perspective of clustering. The structure of the $3/2^-_3$ state at
$E_x=8.56$~MeV has rarely been theoretically discussed, although cluster
structures of highly excited states of $^{11}$B above $E_x=10$~MeV have
been suggested by cluster model calculations~\cite{NISH79}.

For clarification of the cluster
structure, further information on the natural-parity
excitation strengths is necessary. Especially, the isoscalar parts of
the natural-parity transition strengths are a key ingredient because
most of the cluster states are excited by the isoscalar natural-parity
transitions. 

The natural-parity transition strengths have been extensively
examined by means of $\gamma$-decay and $(e,e')$ measurements.
However, such electromagnetic probes are sensitive only to the
transitions relevant to protons.
Therefore, the electric transition strength
carried by protons is generally different from the isoscalar strength
carried by both protons and neutrons. 

The electric $[B(E\lambda)]$ and isoscalar
$[B(E\lambda;IS)]$ transition strengths with a multipolarity of
$\lambda$ are described by
\begin{align*}
B(E\lambda)&=e^2\left\lvert M_p(E\lambda)\right\rvert^2/(2J_i+1),\\
B(E\lambda;IS)&=\left\lvert
M_p(E\lambda)+M_n(E\lambda)\right\rvert^2/(2J_i+1),
\end{align*}
where $J_i$ is the spin of the initial state. $M_p(E\lambda)$ and
$M_n(E\lambda)$ are the proton and neutron transition matrix
elements. $M_p(E\lambda)$ and $M_n(E\lambda)$ are given by
\begin{align*}
M_p(E\lambda)&={\langle}f\lVert
(1-\tau_z)\hat{O}(E\lambda){\rVert}i\rangle/2,\\
M_n(E\lambda)&={\langle}f\lVert
(1+\tau_z)\hat{O}(E\lambda){\rVert}i\rangle/2,
\end{align*}
where $\tau_z$ is the isospin Pauli matrix and $\hat{O}(E\lambda)$ is
the transition operator. $\hat{O}(E\lambda)=r^2$ and $r^2Y_2$ are used
for the monopole and quadrupole transitions,
respectively. 

For light self-conjugate 
nuclei, the isoscalar strengths are deduced from the electric transition
strengths because the neutron transition strengths are similar to the
proton transition strengths due to the approximately conserved charge
symmetry. For the other nuclei, however, the neutron and proton
transition strengths are different each other, and the
isoscalar strengths should be determined by a variety of different
measurements.

One possible method to obtain the isoscalar transition strengths is
to measure the electric transition strengths for mirror nuclei. 
In case of the $A=11$ system, the electric quadrupole strengths
have been measured for low-lying states in the stable $^{11}$B nucleus,
but no electric quadrupole strength is reported for the excited
states in unstable $^{11}$C. For the electric monopole strengths,
no experimental value is known for both the states in $^{11}$B and
$^{11}$C.

Another possible method to obtain the isoscalar transition strengths 
is to measure the hadron scattering \cite{HARA01}.
Hadron scattering at forward angles and at intermediate
energies is a good probe to obtain such excitation strengths
thanks to a good proportionality between the cross sections
and the relevant excitation strengths. Since both the
isoscalar and isovector transitions coherently contribute to the excitation
strengths in $N\,{\neq}\,Z$ nuclei with non-zero ground-state isospin,
isoscalar probes like deuteron or $^4$He are useful to extract the
isoscalar excitation strengths.

In the present work, the isoscalar monopole and quadrupole excitation
strengths in $^{11}$B have been obtained by scrutinizing the data from
the $^{11}$B$(d,d')$ reaction at $E_d=200$~MeV previously measured at
the Research Center for
Nuclear Physics, Osaka University \cite{PAPA04}. The excitation
strengths have been compared with the theoretical values by the
antisymmetrized molecular-dynamics (AMD) and the SM
calculations. The former method has been demonstrated to be useful for
describing the cluster structure of the light nuclei
\cite{ENYO01,ENYO03}. By analyzing the isoscalar excitation strengths,
the cluster structure of the excited states in $^{11}$B are probed.

%%%%%%%%%%%%%%%%%%%%%%%%%%%%%%%%%%%%%%%%%%%%%%%%%%%%%%%%%%%%%%%%%%%%%%%
%\section{Experiment and results}

%The experiment was performed at the Research Center for Nuclear Physics,
%Osaka University, using a 200-MeV deuteron beam.
%The deuteron beam extracted from the ring cyclotron was achromatically
%transported to a self-supporting $^{11}$B target with a thickness of
%16.7 mg/cm$^2$.
%Scattered deuterons were momentum analyzed by the
%high-resolution spectrometer Grand Raiden \cite{MAMO99}. 
%An energy resolution of 150 keV full width at
%half maximum was obtained for the beam intensity of 1--10 nA.

\begin{figure}
\centering
\includegraphics[scale=0.5]{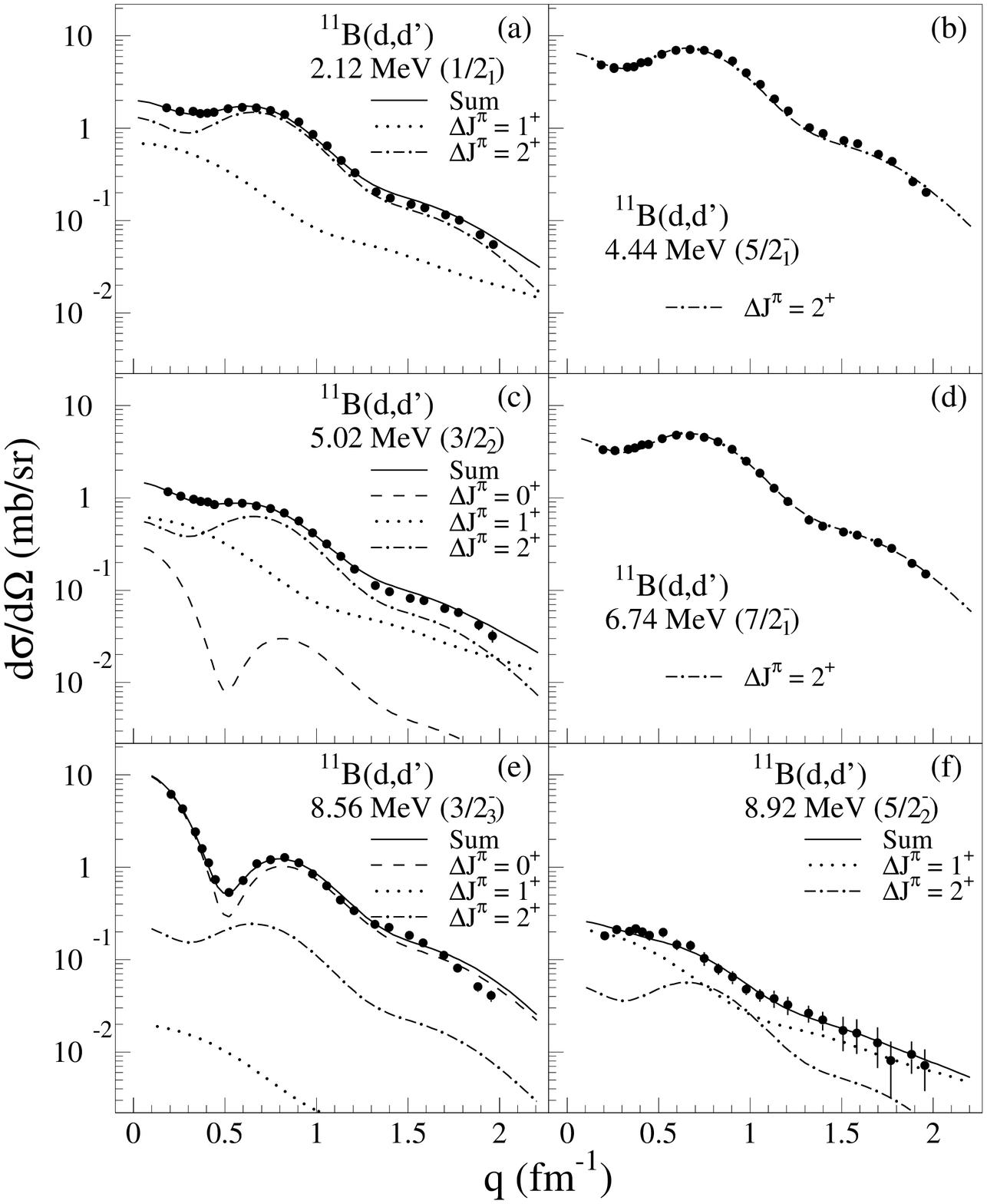}%
\caption{\label{fig:b11dd}
Cross sections for 
the $^{11}$B$(d,d')$ reaction at $E_d=200$~MeV.
The dashed, dotted, and dash-dotted curves show the ${\Delta}J^\pi=0^+$,
$1^+$, and $2^+$ contributions, respectively. The
solid curves are the sums of all the multipole contributions. 
This figure is same with Fig.~4 in Ref.~\cite{PAPA04}.
}
\end{figure}

The $^{11}$B$(d,d')$ cross sections were analyzed by summing up the
cross sections calculated with various multipole transitions since the
spin-parity of the ground state of $^{11}$B is $3/2^-$. 
The cross section for each multipole transition was calculated
in the framework of the macroscopic model in which the transition
potential was obtained in the prescription of the deformed potential
model \cite{HARA01}.
Since the angular distribution of the cross section for each multipole
transition depends on its transferred angular momentum, it is possible to
decompose the cross section into each multipole component by fitting 
the measured angular distribution as shown in Fig.~\ref{fig:b11dd}. 
In the fitting procedure, the multipole contributions
with ${\Delta}J\ge3$ were neglected. 

From the multipole decomposition analysis (MDA), the isoscalar monopole and
quadrupole strengths were extracted. Table~\ref{tab:be2} lists the
obtained $B(E0; IS)$ and $B(E2; IS)$ values together with the $B(E2)$
values taken from Ref.~\cite{AJZE90}.
Systematic uncertainties on $B(E0; IS)$ and $B(E2; IS)$ are mainly
due to errors in the model calculation for the ($d,d'$) reaction. 
The detailed explanation for MDA
has been given in Ref.~\cite{PAPA04}.

Although the $3/2^-$ states are allowed to be excited by the
${\Delta}J^\pi=0^+$,
$1^+$, $2^+$, and $3^+$ transitions, the
${\Delta}J^\pi=0^+$ contribution in exciting the 8.56-MeV $3/2^-_3$
state is found to be extraordinary large.
Since the observed ${\Delta}J^\pi=0^+$ strength is
much stronger than the expected ${\Delta}J^\pi=2^+$ strength, it is
difficult to reliably extract the ${\Delta}J^\pi=2^+$ transition
strength for the $3/2^-_3$ state. 
For the 5.02-MeV $3/2^-_2$ state, the monopole strength is not reliably
extracted because the ${\Delta}J^\pi=0^+$ strength is much weaker than
the other multipole components.

%%%%%%%%%%%%%%%%%%%%%%%%%%%%%%%%%%%%%%%%%%%%
%\section{Discussion}

\begin{figure}
\centering
\includegraphics[scale=0.6]{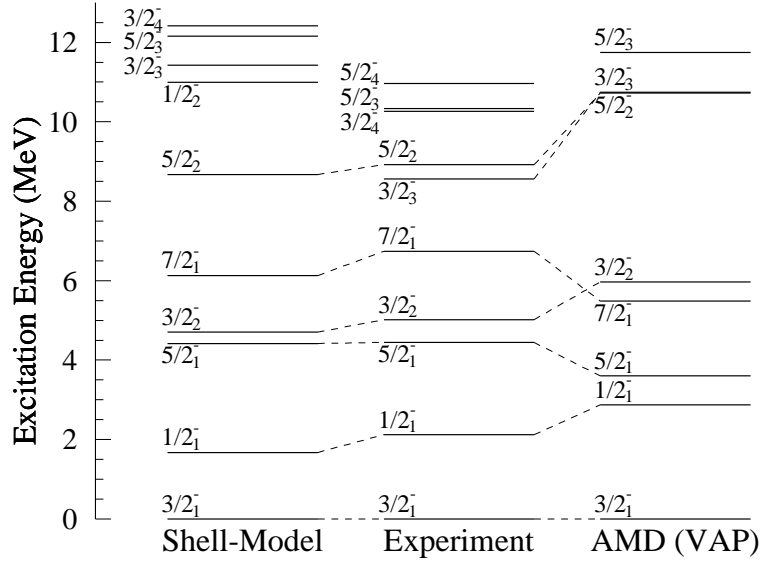}%
\caption{\label{fig:level}
Experimental and theoretical level schemes for the
negative-parity states in $^{11}$B.}
\end{figure}

The SM calculation was performed with the SFO
(Suzuki-Fujimoto-Otsuka) interactions \cite{SUZU03} within the
0--2$\hbar\omega$ configuration space. 
The level schemes for the negative-parity states are
compared with the experiment in Fig.~\ref{fig:level}. 
The harmonic oscillator potential was used to calculate the
single-particle wave functions. 
The oscillator lengths were obtained from the DWIA analysis of the
($^3$He,$t$) and ($p,p'$) reactions \cite{PAPA04}.
Since the quadrupole strengths with the bare charges of
$e_p=1$ and $e_n=0$ were much smaller than the experimental values by a
factor of 2--3, the effective charges were introduced to improve the
theoretical prediction. 
The best-fit results of the
quadrupole strengths were obtained with the effective charges of
$e^{eff}_p=1.24$ and  $e^{eff}_n=0.22$ as tabulated in
Table~\ref{tab:be2}.
The obtained effective charges are slightly smaller than the standard
values of $e^{eff}_p\sim1.3$ and  $e^{eff}_n\sim0.5$ for light stable
nuclei. 

Although the SM calculation reasonably well explains the
experimental $B(E2;IS)$ and $B(E2)$ values for the
low-lying states, the description for the transition properties for
states at $E_x\sim9$~MeV
is not reasonable. For the $5/2^-_2$ state at $E_x=8.92$~MeV, the SM
calculation gives extremely small quadrupole strengths and
underestimates the experimental data although
the spin-flip $M1$ strengths are well described in Ref.~\cite{PAPA04}. 

The SM calculation also failed to describe the observed
$3/2_3^-$ state at $E_x=8.56$~MeV.
Although the $3/2_3^-$ state appears at $E_x=11.4$ MeV in the
SM calculation, its transition properties are completely
different from those of the observed $3/2_3^-$ state.
The predicted $3/2_3^-$ state carries almost no monopole excitation
strength, while the observed $3/2_3^-$ state is dominantly excited
by the ${\Delta}J^\pi=0^+$ transition [see Fig.~\ref{fig:b11dd}(e)].
Hence, the predicted $3/2^-_3$ state is different from the observed
$3/2^-_3$ state.

It is noteworthy to point out the analogies between the $3/2^-_3$ state
in $^{11}$B and the $0^+_2$ state at $E_x=7.65$~MeV in $^{12}$C which
is excited from the ground state with an isoscalar monopole strength of
$B(E0;IS)=121{\pm}9$~fm$^4$ \cite{AJZE90}.
Both the two states locate at the excitation energies near the
$\alpha$-decay threshold, while they are not satisfactorily predicted in
the SM calculation. They carry the large isoscalar monopole strengths
with a similar magnitude.  
Therefore, the observed $3/2^-_3$ state in $^{11}$B is
inferred to have a structure analogous with that of the $0^+_2$ state
in $^{12}$C from the view point of cluster physics. Namely, the $3/2^-_3$
state is expected to have a dilute cluster structure with a $2\alpha+t$
configuration in the same manner as that the $0^+_2$ state in $^{12}$C
has a dilute $3\alpha$ structure.

To examine cluster nature of excited states in $^{11}$B, the
experimental results are compared with the AMD calculation.
The method of the variational calculation after the spin-parity
projection (VAP) was used as described in
Refs.~\cite{ENYO98,ENYO99,ENYO05} where the properties of the excited
states in $^{12}$C and $^{10}$Be were reasonably well explained. The MV1
\cite{ANDO80} and G3RS \cite{YAMA79,TAMA68} interactions were used for
the central and spin-orbit forces in the calculation, respectively. The
adopted interaction parameters of $m=0.62$, $b=h=0.25$, and
$u_I=-u_{II}=2800$~MeV were the same as those in Ref.~\cite{ENYO05}.

The calculated excitation energies, monopole and quadrupole strengths in
$^{11}$B are shown in Fig.~\ref{fig:level} and are listed in
Table~\ref{tab:be2}. In the AMD (VAP) calculations, the $7/2^-_1$ and
$5/2^-_2$ states locate below the $3/2^-_2$ and $3/2^-_3$ states,
respectively. The excitation energies for the $3/2^-_3$ and
$5/2^-_2$ states are higher than the experimental values by about
2~MeV. The predicted level structure is slightly different from the
experimental level scheme of $^{11}$B. However, the excitation strengths
are reasonably well reproduced without introducing any effective
charges. In addition, the large monopole strength for the $3/2^-_3$
state, which is not predicted by SM calculations, is successfully
predicted.

For the $5/2^-_2$ state, the AMD (VAP) calculation reasonably explains
both the quadrupole and spin-flip $M1$ strengths although the
experimental uncertainties for the quadrupole strengths are large.
The calculated wave function for the $5/2^-_2$ state is dominated by a
SM-like component with a mixing of the cluster component, which is
characterized by cluster correlation at a SU(3) limit.
The cluster component provides almost no spin-flip $M1$ strength between
the $3/2^-$ ground state and the $5/2^-_2$ state but enhances the
quadrupole strength, while the SM-like component provides the
significant $M1$ strength but almost no quadrupole strength.
This suggests that both the SM-like and cluster structures 
should coexist in the $5/2^-_2$ state to explain the sizable spin-flip
$M1$ and quadrupole strengths simultaneously, and the cluster correlation
plays a role to enhance the $E2$ strengths for the $5/2^-_2$ state. To
clarify the detailed structure of the $5/2^-_2$ state, the precise
measurement of the $E2$ strength is desired.

The $B(E2)$ value for the $3/2^-_3$ state is predicted to be
0.84~$e^2$fm$^4$, while the reported value from the $(e,e')$ measurement
is as large as $9.4{\pm}0.2$~$e^2$fm$^4$ \cite{KAN75}.
However, this reported value is not reliable because only the $M1$ and
$E2$ transitions were taken into account and the $E0$
transition was neglected in the previous analysis \cite{KAN75,SPAM66}. 
We have analyzed the existing $(e,e')$ data again by taking the
$E0$ and $M1$ transitions into account and by neglecting the $E2$ transition
according to the suggestion from the AMD (VAP) calculation.
As the result, the large $B(E0)$ value of $18.7{\pm}0.7$~$e^2$fm$^4$ has
been obtained. Assuming a simple relation of $M_p(E0)=(Z/N)M_n(E0)$,
the $B(E0;IS)$ value of $90{\pm}3$~fm$^4$ is obtained from this $B(E0)$
value. The large $B(E0)$ and negligibly small $B(E2)$ values are quite
consistent with the present experimental and theoretical results in
Table~\ref{tab:be2}.

Regarding the electromagnetic transition between the $3/2^-_3$ state and
the ground state, it is noteworthy to discuss the result on the
correlation measurement of electron-positron pairs emitted in the
internal pair formation decay \cite{OLNE65}. 
Although it is concluded in Ref.~\cite{OLNE65} that
the $E2$ transition dominates the decay of the $3/2^-_3$ state,
the observed correlation of the electron-positron pairs
is also explained by assuming the mixed $E0$-$M1$ transition.
The $E0$ and $M1$ strengths obtained from the present
$(e,e')$ analysis reasonably account for the correlation observed in
Ref.~\cite{OLNE65}.

In contrast to the large monopole strength for the $3/2^-_3$ state,
the monopole strength for the $3/2^-_2$ state is small. This difference
in the monopole strengths is well explained by the AMD (VAP) calculation.
According to the calculation, the $3/2^-_3$ state has a spatially
well-developed cluster structure with a loosely bound $2{\alpha}+t$
configuration, while the spatial development of the $2{\alpha}+t$
cluster structure in the $3/2^-_2$ state is weak.
Therefore, the large monopole strength for the
$3/2^-_3$ state obtained in the present study is regarded to be
evidence of the developed $2\alpha+t$ cluster structure.

To evaluate the dilution of the density
distribution quantitatively, we introduce a new quantity $D$ which is
defined by
$D=\int_{\frac{\rho(r)}{\rho_0}<\frac{1}{5}}\rho(r)\,d^3r/A$ 
where $A$ is a mass number,
$\rho(r)$ is a matter density, and $\rho_0$ is a normal density which is
chosen to be $\rho_0=0.16$ nucleons/fm$^3$. 
The $D$ value provides a fraction of nucleons
in the low density region where the matter density is lower than 1/5 of
the normal density. Table~\ref{tab:rms} lists the root mean square (rms)
radii and the $D$ values for several states in $^{11}$B and $^{12}$C
estimated by the AMD (VAP) and $\alpha$ condensate-model (ACM)
\cite{FUNA05} wave functions.
The ACM wave function gives the large rms radius and $D$
value for the $0^+_2$ state in $^{12}$C, which attracts broad interest
in view of dilute cluster states. 
Since the AMD framework is a kind of the bound-state approximation,
the AMD calculation tends to underestimate the tail of the density
distribution at a large radius.
Actually, the rms radius and the $D$ value for the $0^+_2$ state
estimated by the AMD (VAP) wave function are smaller than those by the
ACM wave function.
However, they are still extraordinary large compared with those for the
ground state of $^{12}$C.
Although the rms radius and $D$ value for the $3/2^-_3$ state in
$^{11}$B in the AMD (VAP) calculation are also smaller than those for
the $0^+_2$ state in the ACM calculation, they
are as large as those for the $0^+_2$ state calculated by AMD (VAP) and
are significantly larger than those for the ground and $5/2^-_2$ states.
The $5/2^-_2$ and $3/2^-_3$ states locate at the same
excitation energy (see Fig.~\ref{fig:level}),
but the rms radius and $D$ value of the $5/2^-_2$ state are much smaller
than those of the $3/2^-_3$ state. On the basis of these arguments, it
is natural to suggest that the $3/2^-_3$ state in $^{11}$B has a dilute
structure. 

The $3^-_1$ state in $^{12}$C also possesses the
large rms radius. According to the AMD (VAP) calculation, however, the
$3^-_1$ state has a developed $3\alpha$ cluster structure with a
triangle configuration, and its structure is different from the 
dilute-gas-like structures of the $0^+_2$ state in $^{12}$C and the
$3/2^-_3$ state in $^{11}$B where the constituent clusters are freely
moving \cite{ENYO06}. 
%Because of the developed triangle $3\alpha$cluster structure, the $D$
%value of the $3^-_1$ state is smaller than those for the $0^+_2$ and
%$3/2^-_3$ states in spite of the large rms radius.

\begin{figure}
\includegraphics[scale=0.6]{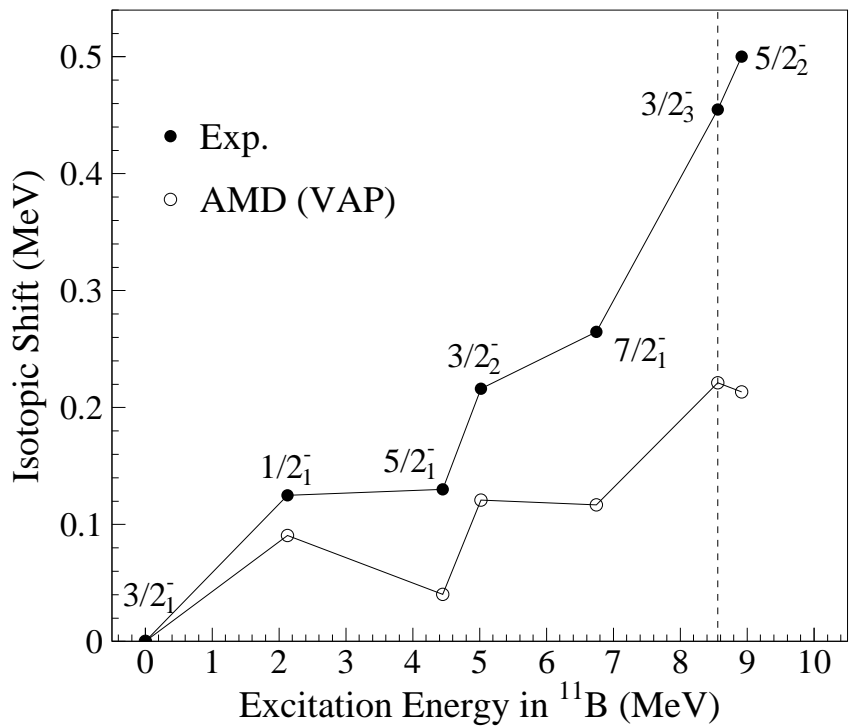}%
\caption{\label{fig:iso}
Isotopic shifts in the excitation energies for the negative-parity
states in the $^{11}$B-$^{11}$C mirror system. The solid circles show
the experimental values, and the open circles are the theoretical
predictions by the AMD (VAP) calculation. The excitation energy of the
$3/2^-_3$ state is shown by the dashed line. 
}
\end{figure}

A manifestation of the dilute structure of the $3/2^-_3$ state may
appear in the
isotopic shift in the excitation energies since the spatial increase in
size of the proton distribution causes the reduction of the Coulomb
energy, although the isotopic shift is caused not only by the
spatial expansion of the proton distribution but also by the nuclear
deformation and single-particle excitation.
Fig.~\ref{fig:iso} shows the experimental and theoretical values
of the isotopic shifts for the negative-parity states in the
$^{11}$B-$^{11}$C mirror system as a function of the experimental
excitation energies in $^{11}$B.
The theoretical predictions are obtained by the AMD (VAP) calculation
where the energies for the excited states in $^{11}$C are calculated by
using the mirror-symmetric AMD wave functions.

The isotopic shifts for the $3/2^-_3$ and $5/2^-_2$ states are much
larger than those for the $3/2^-_2$ and $7/2^-_1$ states.
The $5/2^-_2$ state, which is predicted to have a compact structure,
is considered to exhibit a large isotopic shift due to the
spin-flip single-particle excitation. On the other hand, the large
isotopic shift for the $3/2^-_3$ state is inferred to be a reflection of
the dilute structure discussed in the present paper.

Since the AMD calculation tends to underestimate the long tail of the
wave function at a large radius, it is expected that the calculation is
not good enough to quantitatively reproduce the isotopic shifts.
Actually, the theoretical values of the isotopic shift are
systematically smaller than the experimental values by a factor of about
2. However, the AMD (VAP) calculation, which suggests the dilute
structure of the $3/2^-_3$ state, well reproduces a general trend of the
isotopic shifts and explains state-by-state dependence of the isotopic
shifts. 
%The noticeable increase of the isotopic shift between the $7/2^-_1$ and
%$3/2^-_3$ states is theoretically considered to be a reflection of the
%dilute structure of the $3/2^-_3$ state.

%%%%%%%%%%%%%%%%%%%%%%%%%%%%%%%%%%%%%%%%%%%%

%%%%%%%%%%%%%%%%%%%%%%%%%%%%%%%%%%%%%%%%%%%%%%%%%%%%%%%%%%%%%%%%%%%%%%%
%\section{Summary}

In summary, 
the isoscalar monopole and quadrupole excitation strengths for the
low-lying states in $^{11}$B were determined by measuring 
the $^{11}$B($d,d'$) reaction.
The obtained excitation strengths were compared
with the SM calculation using the SFO interaction and with the
AMD (VAP) calculation. 
The $3/2^-_3$ state is excited with a strong monopole strength, and 
is inferred to have a $2\alpha+t$ cluster wave function in analogy
with the $0^+_2$ state in $^{12}$C which is known to have a
dilute-gas-like $3\alpha$ cluster structure.
From the analysis of the monopole excitation strengths with the AMD
calculations, the $3/2^-_3$ state is suggested to have a
loosely bound $2\alpha+t$ cluster structure with a dilute density.

\begin{ack}
The authors would like to thank Prof.~H.~Horiuchi,
Prof.~Toshio~Suzuki, Prof.~T.~Otsuka, Dr.~Y.~Funaki, and Dr.~S.~Fujii
for valuable discussions.
The authors acknowledge the effort
of the RCNP cyclotron crew for providing the stable and clean
beam. Numerical calculations are partially done with the supercomputer
in KEK and YITP (Kyoto). This research was supported in part by the
Grant-in-Aid for Scientific Research No. 15740136 and 17740132 from the
Japan Ministry of Education, Sports, Culture, Science, and Technology.
\end{ack}

% The Appendices part is started with the command \appendix;
% appendix sections are then done as normal sections
% \appendix

% \section{}
% \label{}

\newpage

\begin{table*}
\caption{\label{tab:be2}
Measured monopole and quadrupole strengths for the excited states in
$^{11}$B compared with the theoretical predictions by the
SM \cite{SUZU03} and AMD (VAP) calculations. 
%In the shell-model
%calculation, the effective charges of $e^{eff}_p=1.24$ and
% $e^{eff}_n=0.22$ are used.
}
\centering
{\scriptsize
\tabcolsep = 0.55 mm
%\begin{ruledtabular}
\begin{tabular}{cccccccccccccc}
\hline
& &\multicolumn{2}{c}{Present}& &
Ref.~\cite{AJZE90}& &
\multicolumn{2}{c}{Shell-model}& &
\multicolumn{4}{c}{AMD (VAP)}\\ 
\cline{3-4} \cline{6-6}
\cline{8-9} \cline{11-14}
$E_x$ & $J^\pi$ & 
$B(E0; IS)$ & $B(E2; IS)$ & &
$B(E2)$ & &
$B(E2; IS)$ & $B(E2)$ & &
$B(E0; IS)$ & $B(E0)$ & $B(E2; IS)$ & $B(E2)$ \\
(MeV)& & 
(fm$^4$) & (fm$^4$) & &
($e^2$fm$^4$) & &
(fm$^4$) & ($e^2$fm$^4$) & &
(fm$^4$) & ($e^2$fm$^4$) & (fm$^4$) & ($e^2$fm$^4$) \\
\hline
2.12 & $1/2^-_1$ &
          & $11\pm2$    & & $2.6\pm0.4$ & &
 12.0     &  1.8        & &
          &             &
12.3      &  2.3 \\
4.44 & $5/2^-_1$ &
          & $56\pm6$    & & $21\pm6$    & &
 49.5     &  16.5       & &
          &             &
 66.5     & 19.2 \\
5.02 & $3/2^-_2$ &
 $<9$     & $4.7\pm1.5$ & & $<1.3$      & &
 14.2     &  1.7        & &
  7       &  1.3        &
 2.3      &  0.02 \\
6.74 & $7/2^-_1$ &
          & $38\pm4$    & & $3.7\pm0.9$ & &
 42.9     &  4.4        & &
          &             &
 34.4     &  3.6 \\
8.56 & $3/2^-_3$ &
$96\pm16$ & $<6$        & &             & &
 0.012*   &  0.15*      & &
 94       &  19         &
 5.3      &  0.84 \\
8.92 & $5/2^-_2$ &
          & $0.4\pm0.3$ & & $1.6\pm1.2$ & &
 0.012    &  0.014      & &
          &             &
 0.66     &  0.15 \\
\hline
\end{tabular}
%\end{ruledtabular}
}
\flushleft{\scriptsize *The $3/2^-_3$ state predicted by the SM
 calculation is different from the observed $3/2^-_3$ state (see text).}
\end{table*}

\begin{table}
\caption{\label{tab:rms}
Root mean square radii $(\sqrt{{\langle}r^2\rangle})$ and fractions of
the nucleon numbers in low density region with $\rho/\rho_0<1/5$ $(D)$
estimated by the AMD (VAP) and ACM \cite{FUNA05} wave functions.
The AMD (VAP) calculation for $^{12}$C are performed by using the same
interaction as that in Ref.~\cite{ENYO98}.
}
%{\footnotesize
%\begin{ruledtabular}
\centering
\begin{tabular}{lccccc}
\hline
&\multicolumn{2}{c}{AMD (VAP)} & &\multicolumn{2}{c}{ACM}\\
\cline{2-3} \cline{5-6}
 & $\sqrt{{\langle}r^2\rangle}$ (fm) & $D$ & &
$\sqrt{{\langle}r^2\rangle}$ (fm) & $D$ \\ \hline
$^{11}$B $3/2^-_1$        & 2.5 & 0.29 & &     &      \\
$^{11}$B $3/2^-_3$        & 3.0 & 0.42 & &     &      \\
$^{11}$B $5/2^-_2$        & 2.6 & 0.24 & &     &      \\
$^{12}$C $0^+_1$          & 2.5 & 0.21 & & 2.4 & 0.18 \\
$^{12}$C $0^+_2$          & 3.3 & 0.45 & & 3.8 & 0.68 \\
$^{12}$C $3^-_1$          & 3.1 & 0.38 & &     &      \\
\hline
\end{tabular}
%\end{ruledtabular}
%}
\end{table}


\begin{thebibliography}{00}

% \bibitem{label}
% Text of bibliographic item

% notes:
% \bibitem{label} \note

% subbibitems:
% \begin{subbibitems}{label}
% \bibitem{label1}
% \bibitem{label2}
% If there is a note, it should come last:
% \bibitem{label3} \note
% \end{subbibitems}


\bibitem{IKED68}
  K.~Ikeda et al.,
  Prog. Theor. Phys. Suppl. Extra Number (1968) 464.
\bibitem{MORI56}
  H.~Morinaga, Phys. Rev. 101 (1956) 254.
\bibitem{FUJI80}
  Y.~Fujiwara et al.,
  Prog. Theor. Phys. Suppl. 68 (1980) 29.
\bibitem{UEGA77}
  E.~Uegaki et al.,
  Prog. Theor. Phys. 57 (1977) 1262.
\bibitem{UEGA79}
  E.~Uegaki et al.,
  Prog. Theor. Phys. 62 (1979) 1621.
\bibitem{KAMI81}
  M.~Kamimura,
  Nucl. Phys. A 351 (1981) 456.
\bibitem{TOHS01}
  A.~Tohsaki et al.,
  Phys. Rev. Lett. 87 (2001) 192501.
\bibitem{FUNA03}
  Y.~Funaki et al.,
  Phys. Rev. C 67 (2003) 051306.
\bibitem{FUNA05}
  Y.~Funaki {\it et al.},
  Eur. Phys. J. A 24 (2005) 321.
\bibitem{YAMA05}
  T.~Yamada et al.,
  Eur. Phys. J. A 26 (2005) 185.
\bibitem{MATS04}
  H.~Matsumura {\it et al.},
  Nucl. Phys. A 739 (2004) 238.
\bibitem{YAMA04}
  T.~Yamada et al.,
  Phys. Rev. C 69 (2004) 024309.
\bibitem{PAPA04}
  T. Kawabata et al.,
  Phys. Rev. C 70 (2004) 034318.
\bibitem{FUJI04}
  Y. Fujita et al.,
  Phys. Rev. C 70 (2004) 011306.
\bibitem{HANN03}
  V.~M.~Hannen {\it et al.},
  Phys. Rev. C 67 (2003) 054320.
\bibitem{NISH79}
  H. Nishioka et al., 
  Prog. Theor. Phys. 62 (1979) 424.
\bibitem{HARA01}
  M.~N.~Harakeh and A.~van~der~Woude, {\it Giant Resonances}
  (Oxford Univ. Press, 2001).
\bibitem{ENYO01}
  Y.~Kanada-En'yo et al.,
  Prog. Theor. Phys. Suppl. 142 (2001) 205.
\bibitem{ENYO03}
  Y.~Kanada-En'yo et al.,
  C. R. Physique 4 (2003) 497.
\bibitem{AJZE90}
  F. Ajzenberg-Selove,
  Nucl. Phys. A 506 (1990) 1.
\bibitem{SUZU03}
  T.~Suzuki et al.,
  Phys. Rev. C 67 (2003) 044302.
\bibitem{ENYO98}
  Y.~Kanada-En'yo,
  Phys. Rev. Lett. 81 (1998) 5291.
\bibitem{ENYO99}
  Y.~Kanada-En'yo et al.,
  Phys. Rev. C 60 (1999) 064304.
\bibitem{ENYO05}
  Y.~Kanada-En'yo,
  RIKEN Accel. Prog. Rep. 39 (2006) 16.
\bibitem{ANDO80}
  T.~Ando et al.,
  Prog. Theor. Phys. 64 (1980) 1608.
\bibitem{YAMA79}
  N.~Yamaguchi et al.,
  Prog. Theor. Phys. 62 (1979) 1018.
\bibitem{TAMA68}
  R.~Tamagaki,
  Prog. Theor. Phys. 39 (1968) 91.
\bibitem{KAN75}
  P.~T.~Kan et al.,
  Phys. Rev. C 11 (1975) 323.
\bibitem{SPAM66}
  E.~Spamer,
  Z. Phys. 191 (1966) 24.
\bibitem{OLNE65}
  J.~W.~Olness {\it et al.},
  Phys. Rev. 139 (1965) B512.
\bibitem{ENYO06}
  Y.~Kanada-En'yo, nucl-th/0605047.
\end{thebibliography}
\end{document}